\documentclass[12pt,english]{article}
\usepackage{avant}
\usepackage[T1]{fontenc}
\usepackage[latin9]{inputenc}
\usepackage{geometry}
\usepackage{babel}
\usepackage{array}
\usepackage{float}
\usepackage{multirow}
\usepackage{amsmath}
\usepackage{amssymb}
\usepackage{stmaryrd}
\usepackage[authoryear]{natbib}
\usepackage[unicode=true,
 bookmarks=false,
 breaklinks=false,pdfborder={0 0 1},colorlinks=false]
 {hyperref}
\usepackage{breakurl}

\makeatletter

\providecommand{\tabularnewline}{\\}
\floatstyle{ruled}
\newfloat{algorithm}{tbp}{loa}
\providecommand{\algorithmname}{Algorithm}
\floatname{algorithm}{\protect\algorithmname}

\usepackage{amsthm}\usepackage{bm}\usepackage{graphicx}\usepackage{rotating}\usepackage{array}\usepackage{booktabs}\usepackage{color}\usepackage{algorithmic}
\usepackage{algorithm}\usepackage{multirow}\usepackage{enumerate}\usepackage{graphicx}\usepackage{caption}\usepackage{subcaption}

\floatstyle{ruled}
\newfloat{algorithm}{tbp}{loa}
\providecommand{\algorithmname}{Algorithm}
\floatname{algorithm}{\protect\algorithmname}

\usepackage{vmargin}
\setmarginsrb{1.0in}{0.5in}{1.0in}{1.5in}{0.5in}{0.5in}{0.5in}{0.5in}

\providecommand{\tabularnewline}{\\}

\@ifundefined{definecolor}{color}{}
\RequirePackage[displaymath]{lineno}

\@ifundefined{definecolor}{color}{}
\usepackage{amsfonts}\usepackage{latexsym}

\usepackage{mathrsfs}\usepackage{amsthm}\usepackage{latexsym}\usepackage{booktabs}\@ifundefined{definecolor}{color}{}

\usepackage{times}\@ifundefined{definecolor}{color}{}

\numberwithin{equation}{section}
\numberwithin{figure}{section}

\newtheorem{thm}{Theorem}
\newtheorem{lem}{Lemma}
\newtheorem{cy}{Corollary}
\newtheorem{pn}{Proposition}

\newtheorem{defi}{Definition}

 \newcommand{\T}{^{\mbox{\tiny {\sf T}}}}

\usepackage{babel}

\makeatother

\begin{document}
{\setlength{\baselineskip}{1.5\baselineskip} 

\global\long\def\mbA{\mathbf{A}}
 \global\long\def\mbB{\mathbf{B}}
 \global\long\def\mbD{\mathbf{D}}
 \global\long\def\mbN{\mathbf{N}}
 \global\long\def\mbV{\mathbf{V}}
 \global\long\def\mbv{\mathbf{v}}
 \global\long\def\mbX{\mathbf{X}}
 \global\long\def\mbY{\mathbf{Y}}
 \global\long\def\mbU{\mathbf{U}}
 \global\long\def\mbW{\mathbf{W}}
 \global\long\def\mbP{\mathbf{P}}
 \global\long\def\mbQ{\mathbf{Q}}
 \global\long\def\mbO{\mathbf{O}}
 \global\long\def\mbM{\mathbf{M}}
 \global\long\def\mbR{\mathbf{R}}
 \global\long\def\mbS{\mathbf{S}}
 \global\long\def\mbG{\mathbf{G}}
 \global\long\def\mbI{\mathbf{I}}
 \global\long\def\hatmbG{\widehat{\mbG}}
 \global\long\def\mbg{\mathbf{g}}
 \global\long\def\hatmbg{\widehat{\mbg}}
 \global\long\def\hatmbM{\widehat{\mbM}}
 \global\long\def\hatmbW{\widehat{\mbW}}
 \global\long\def\hatmbU{\widehat{\mbU}}
 \global\long\def\mbw{\mathbf{w}}
 \global\long\def\hatmbw{\widehat{\mbw}}
 \global\long\def\bolGamma{\boldsymbol{\Gamma}}
 \global\long\def\hatbolGamma{\widehat{\bolGamma}}
 \global\long\def\bolOmega{\boldsymbol{\Omega}}
 \global\long\def\bolPhi{\boldsymbol{\Phi}}
 \global\long\def\mbbR{\mathbb{R}}
 \global\long\def\mbbS{\mathbb{S}}
 \global\long\def\calE{\mathcal{E}}
 \global\long\def\boltheta{\boldsymbol{\theta}}
 \global\long\def\boleta{\boldsymbol{\eta}}
 \global\long\def\bolphi{\boldsymbol{\phi}}
 \global\long\def\bolpsi{\boldsymbol{\psi}}
 \global\long\def\hatboltheta{\widehat{\boltheta}}
 \global\long\def\bolbeta{\boldsymbol{\beta}}
 \global\long\def\hatbolbeta{\widehat{\bolbeta}}
 \global\long\def\hatboleta{\widehat{\boleta}}
 \global\long\def\hatbolpsi{\widehat{\bolpsi}}
 \global\long\def\hatbolOmega{\widehat{\bolOmega}}
 \global\long\def\bolvarepsilon{\boldsymbol{\varepsilon}}
 \global\long\def\bolalpha{\boldsymbol{\alpha}}
 \global\long\def\bolTheta{\boldsymbol{\Theta}}
 \global\long\def\bolSigma{\boldsymbol{\Sigma}}
 \global\long\def\hatbolSigma{\widehat{\bolSigma}}
 \global\long\def\calR{{\cal R}}
 \global\long\def\calRp{\calR^{\perp}}
 \global\long\def\calG{\mathcal{G}}
 \global\long\def\calU{\mathcal{U}}

\global\long\def\vecc{\mathrm{vec}}
 \global\long\def\Prob{\mathrm{Pr}}
 \global\long\def\E{\mathrm{E}}
 \global\long\def\Cov{\mathrm{cov}}
 \global\long\def\Corr{\mathrm{corr}}
 \global\long\def\F{\mathrm{F}}
 \global\long\def\J{\mathrm{J}}
 \global\long\def\I{\mathcal{I}_{n}}
 \global\long\def\II{\mathcal{I}_{n}^{\mathrm{1D}}}
 \global\long\def\Jn{\mathrm{J}_{n}}
 \global\long\def\Ln{\ell_{n}}
 \global\long\def\Var{\mathrm{var}}
 \global\long\def\dimension{\mathrm{dim}}
 \global\long\def\spn{\mathrm{span}}
 \global\long\def\vech{\mathrm{vech}}

\global\long\def\Env{\mathrm{Env}}
 \global\long\def\tr{\mathrm{trace}}
 \global\long\def\dg{\mathrm{diag}}
 \global\long\def\asyVar{\mathrm{avar}}

\global\long\def\MenvU{\calE_{\mbM}(\mbU)}

\title{{Model-free Envelope Dimension Selection}}
\author{{Xin Zhang}%
\thanks{{\small{Xin Zhang is Assistant Professor, Department of Statistics, Florida State University, Tallahassee, FL, 32306 (Email: henry@stat.fsu.edu).}}%
} \ {and\ Qing Mai}%
\thanks{{\small {Qing Mai is Assistant Professor, Department of Statistics, Florida State University, Tallahassee, FL, 32306 (Email: mai@stat.fsu.edu). }}%
} 
}
\date{}
\maketitle

\begin{abstract}
An envelope is a targeted dimension reduction subspace for simultaneously achieving dimension reduction and improving parameter estimation efficiency. While many envelope methods have been proposed in recent years, all envelope methods hinge on the knowledge of a key hyperparameter, the structural dimension of the envelope. How to estimate the envelope dimension consistently is of substantial interest from both theoretical and practical aspects. Moreover, very recent advances in the literature have generalized envelope as a model-free method, which makes selecting the envelope dimension even more challenging. Likelihood-based approaches such as information criteria and likelihood-ratio tests either cannot be directly applied or have no theoretical justification. To address this critical issue of dimension selection, we propose two unified approaches -- called FG and 1D selections -- for determining the envelope dimension that can be applied to any envelope models and methods. The two model-free selection approaches are based on the two different envelope optimization procedures: the full Grassmannian (FG) optimization and the 1D algorithm \citep{CookZhang2015algorithm}, and are shown to be capable of correctly identifying the structural dimension with a probability tending to 1 under mild moment conditions as the sample size increases. While the FG selection unifies and generalizes the BIC and modified BIC approaches that existing in the literature, and hence provides the theoretical justification of them under weak moment condition and model-free context, the 1D selection is computationally more stable and efficient in finite sample. Extensive simulations and a real data analysis demonstrate the superb performance of our proposals.\\
\end{abstract}


\section{Introduction}
Envelope methods provide means to achieve sufficient dimension reduction and estimation efficiency on a wide range of multivariate statistics problems. The first envelope method was introduced by \citet{Cook2010envelope} in multivariate linear regression to gain efficiency in parameter estimation. Various types of envelope models have been further proposed in multivariate linear regression \citep[etc.]{SuCook2011partial,Cook2013PLS,CookZhang2015simultaneous,CFZ2015RRE}.
More recently, \citet{CookZhang2015foundation} proposed a new definition and framework of envelope that adapted envelope methods to any multivariate parameter estimation procedure. Envelope methods now can be constructed in the model-free context, and are no longer restricted to likelihood-based estimation or stringent regression model assumptions. This greatly facilitates further adaptations of envelope methods to many potential fields such as tensor decomposition and regression with neuroimaging applications \citep{li2016tensor, zhang2016tensor}, Aster models for life history analysis \citep{geyer2007aster, eck2015application}, etc.

All envelope methods rely on the knowledge of the envelope dimension. However, selecting envelope dimension is a theoretically challenging but crucial issue that becomes a severe nag in applications. Even for likelihood-based envelope methods, where information criteria and likelihood-ratio tests are widely used, no theoretical justification is known when the likelihood is mis-specified. To the best of our knowledge, all existing envelope dimension selection procedures in the literature fall into two categories -- either (1) theoretically justified procedures that relying on strong model and distributional assumptions, or, (2) selection procedures based on heuristics such as cross-validation and heuristic information criteria. For example, \citet{Schott2013} provided some pioneering results for likelihood-ratio tests,  \citep{CookZhang2015simultaneous} developed a sequential asymptotic $\chi^2$-test based on rank estimation from \citet{bura2003rank}, and \citet{cook2013scaled} have shown model selection consistency  of BIC under their scaled envelope model and normally distributed errors. All such procedures require the linear model assumption and normality assumptions on either the error or even on the joint distribution of $(\mbY,\mbX)$. It is thus difficult to generalize such approaches to the model-free context and to justify such approaches without normality assumptions. 
On the other hand, information criteria such as AIC \citep{aic} and BIC \citep{bic} are widely used in envelope literature ever since the first paper in envelope \citep{Cook2010envelope}. More recently, \citet{li2016tensor} proposed a modified BIC criterion for the more complicated tensor envelope regression models to estimate the dimension of tensor envelopes on each mode of the tensor. Unfortunately, there is no theoretical justification for BIC or modified BIC in envelope models while the normal assumption or the model assumption is violated. Specifically, without the normality assumption, the envelope estimator is still applicable and is $\sqrt{n}$-consistent estimator for the parameter of interest if we know the true dimension of the envelope, but there is no theory or method available (to the best of our knowledge) for selecting the envelope dimension consistently without relying on the normality assumption or the likelihood. One motivation of this paper is to formally address the theoretical challenges in envelope dimension selection without requiring distributional or model assumptions. 

In this paper, we propose two unified and model-free envelope dimension selection procedures that are applicable to any envelope methods, either model-based or model-free, and suitable for any envelope estimation, either likelihood-based or moment-based. 
Consistency in selecting the envelope dimension is established for both procedures under mild moment conditions and without requiring any particular models.
The first one is called the FG procedure, based on fully optimizing the envelope objective function over a sequence of Grassmannians with increasing dimensions. 
The FG procedure is closely related to the BIC and is shown to include the BIC and the modified BIC \citep{li2016tensor} as special cases. Thus it provides solid theoretical justifications for the popular use of BIC in envelope dimension selection under non-normality and potential model mis-specifications.

From recent developments in envelope algorithms \citep[c.f. 1D algorithm]{CookZhang2015algorithm}, sequentially optimizing a series of objective functions over one-dimensional Grassmannians  can lead to faster, more accurate and stable envelope estimation.
Moreover, because the FG envelope estimation can not guarantee ``nested'' envelope subspace estimates with increasing dimensions, the sequentially nested 1D envelope subspace estimates become even more desirable for its computational simplicity and stability.
We then consider adopt the 1D envelope estimation into envelope dimension selection.
However, as one of our interesting theoretical findings, simply plugging in the 1D envelope estimators into the FG criterion (or BIC/modified BIC) will not guarantee consistency in selecting envelope dimensions.
We thus proposed a new 1D criterion and established consistency in envelope dimension selection with this new criterion.

The contributions of this paper are multi-fold. First of all, ever since the introduction of envelope methods \citep{Cook2010envelope}, there lacks a theoretically well justified approach to selecting its structural dimension in practice. Although \citep{Cook2010envelope} suggested that an information criterion like AIC or BIC may be used to select the structural dimension, no theoretical results were presented to show that such an approach leads to consistent selection if the normality assumption is dropped. In the later papers, BIC has also been applied or modified \citep[e.g.]{li2016tensor} as a working method to select the dimension beyond linear models, while no study exists on the consistency of the BIC type selection. Our paper closes these theoretical gaps for the first time in this research area. Our results complement the existing papers on envelope methods by providing theoretical support to their data analysis. Our studies overcome some major difficulties since we do not rely on any likelihood or model assumptions. Now all the moment-based and the model-free envelope methods (and even future envelope methods) are finally completed with a properly justified model selection criterion.
Secondly, while the moment-based, model-free envelope estimation \citep{CookZhang2015foundation} is essentially a two-stage projection pursuit multivariate parameter estimation relying on a generic objective function of envelope basis, our new formulation in Section~\ref{subsec:quasi} offers a way of viewing model-free envelope estimation as an alternative quasi-likelihood approach involving a key matrix $\mbM$, a parameter of interest $\boltheta$, and a feasible parametrization set $\mathcal{A}_k$ for optimization.
This connection greatly deepens our understanding of model-free envelope methods. It shows that even when no likelihood function is available, we can construct a quasi-likelihood based on methods of moments. We expect that this connection will also facilitate the construction of envelope methods in future research, especially when a likelihood function is not available.
Thirdly, the FG and 1D selection criteria proposed in this paper are tied to the estimation methods in the sense that the FG criterion must be applied with the FG estimator and the 1D criterion must be applied with the 1D estimator. Plugging in an arbitrary root-n consistent estimator into either criteria will generally not guarantee consistency in envelope dimension selection. The link between the estimation methods and selection criteria offers a crucial guidance in practice.

It is also worth mentioning that there have been many methods for determining the dimension of a sufficient dimension reduction subspace \citep[for example]{Zeng2008,ZouChen2012BIC,Ma2015,zhu2006sliced,cook2004determining,
schott1994determining, zhu2016dimensionality, zhu2010dimension}, but the envelope dimension selection problem is very different and arguably more difficult in two aspects. 
First, sufficient dimension reduction methods are restricted to regression problems, whereas envelope methods can be applied to any multivariate parameter estimation. Our work provides a unified approach to select the structure dimension of envelopes under its full generality. 
Secondly, many sufficient dimension reduction methods can be formulated as a generalized eigenvalue problem where the dimension of interest is the rank of some kernel matrix. 
For envelopes, this is not so straightforward, as the envelopes are usually estimated from Grassmannian optimization where no analytic solution can be derived. This is also a part of the reason why we need two  different criteria for different envelope optimizing procedures.
BIC-type criteria have already been used extensively, with proven selection consistency, in the dimension determination problems for sufficient dimension reduction \citep{zhu2006sliced, zhu2010dimension}. While the log-likelihood term in those BIC-type criteria can usually be expressed explicitly as a function of eigenvalues (e.g. equation (10) in \citet{zhu2006sliced}), or modified as the ratio of sums of squared eigenvalues (equation (6.1) in \citet{zhu2010dimension}), the envelope objective function can not be further simplified to derive its asymptotic properties.
Hence, studies on the envelop methods such as in this paper requires much more efforts in the technical proofs.

\section{Review of envelopes and the 1D algorithm}\label{sec:review}

We first review the definitions of reducing subspace and envelope. Besides being the basis for envelope methods, the concept of reducing subspace is also commonly used in functional analysis \citep{Conway1990}, but the notion of ``reduction'' differs from the usual understanding in statistics. 

\begin{defi}\label{def: reducing subspace} (Reducing Subspace) A subspace $\calR\subseteq\mbbR^{p}$
is said to be a reducing subspace of $\mbM\in\mbbR^{p\times p}$ if
$\calR$ decomposes $\mbM$ as $\mbM=\mbP_{\calR}\mbM\mbP_{\calR}+\mbQ_{\calR}\mbM\mbQ_{\calR}$,
where $\mbP_{\calR}$ is the projection matrix onto $\calR$ and $\mbQ_{\calR}=\mbI_{p}-\mbP_{\calR}$
is the projection onto $\calR^{\perp}$. If $\calR$ is a reducing
subspace of $\mbM$, we say that $\calR$ reduces $\mbM$.
\end{defi}
\begin{defi}\label{def: CLC env} (Envelope) The $\mbM$-envelope of $\spn(\mbU)$, denoted by $\calE_{\mbM}(\mbU)$, is the intersection of all reducing subspaces of $\mbM>0$ that contain $\spn(\mbU)$. 
\end{defi}

It can be shown that $\MenvU$ is unique and always exists. The dimension of $\calE_{\mbM}(\mbU)$, denoted by $u$, $0\leq u\leq p$, is important for all envelope methods. A smaller $u$ usually indicates more efficiency gain can be achieved by taking the advantage of envelope structures. 
%Let $\bolGamma\in\mbbR^{p\times u}$ be a semi-orthogonal basis matrix of $\calE_{\bolSigma}(\bolbeta)$.

To see the advantages of envelopes, consider the classical multivariate linear model as an example, 
\begin{equation}
\mbY_{i}=\bolbeta\mbX_{i}+\bolvarepsilon_{i},\quad i=1,\dots,n,\label{mlm}
\end{equation}
where $\mbY_i\in\mbbR^{p\times1}$ is the multivariate response, $\bolvarepsilon_i\sim N(0_p,\bolSigma)$ is independent of $\mbX_i\in\mbbR^q$. To estimate $\bolbeta\in \mathbb{R}^{p\times q}$, \citet{Cook2010envelope} seeks the envelope $\calE_{\bolSigma}(\bolbeta)\subseteq\mbbR^p$ (c.f. Definition~\ref{def: CLC env}). Let $\bolGamma\in\mbbR^{p\times u}$ be a semi-orthogonal basis matrix of $\calE_{\bolSigma}(\bolbeta)$, whose orthogonal completion is $\bolGamma_0\in\mbbR^{p\times (p-u)}$. 
The definition of $\calE_{\bolSigma}(\bolbeta)$ has two implications: 
(1) $\bolGamma\T \mbY$ contains all the information about $\bolbeta$ because $\bolbeta$ resides in $\calE_{\bolSigma}(\bolbeta)$; (2) $\bolGamma\T\mbY$ is independent of $\bolGamma_{0}\T\mbY$ given $\mbX$ because by Definition~\ref{def: reducing subspace}, we can write $\bolSigma=\bolGamma\bolGamma\T\bolSigma\bolGamma\bolGamma\T+\bolGamma_0\bolGamma_0\T\bolSigma\bolGamma_0\bolGamma_0\T=\bolGamma\bolOmega\bolGamma\T+\bolGamma_0\bolOmega_0\bolGamma_0\T$ for some $\bolOmega$ and $\bolOmega_0$. Hence, we can safely reduce immaterial variability in the data by eliminating $\bolGamma_{0}\T\mbY$. Consequently, the envelope estimator promotes efficiency in estimation.
% Under the same model of \eqref{mlm}, \citet{Cook2013PLS} studied the predictor envelope $\calE_{\bolSigma_{\mbX}}(\bolbeta\T)\subseteq\mbbR^{d}$, which is the smallest reducing subspace of the covariance of $\mbX$, $\bolSigma_{\mbX}$, that contains $\spn(\bolbeta\T)$. which is associated with the dimension reduction on predictors instead of responses.

We emphasize that the application of envelopes do not rely on the regression model \eqref{mlm}.  Definition \ref{def: CLC env} is generic and only involves two matrices $\mbM$ and $\mbU$. In a general statistical estimation problem of some  parameter vector $\boltheta\in\mbbR^p$, \citet{CookZhang2015foundation} generalized the notion of envelopes as a way to improve some ``standard'' existing $\sqrt{n}$-consistent estimator $\hatboltheta$. In such general cases where the likelihood function need not be known, they proposed to construct the envelope $\MenvU$ with $\mbU=\boltheta\boltheta\T$ and $\mbM$ being the asymptotic covariance of $\hatboltheta$. To obtain a semi-orthogonal basis matrix estimate for the envelope, $\MenvU=\calE_{\mbM}(\boltheta)$, we solve for $\hatbolGamma\in\mbbR^{p\times u}$ that minimizes
 the generic moment-based objective function:
\begin{equation}
\Jn(\bolGamma)=\log\mid\bolGamma\T\hatmbM\bolGamma\mid+\log\mid\bolGamma\T(\hatmbM+\hatmbU)^{-1}\bolGamma\mid.\label{obj_sample}
\end{equation}
After obtaining $\hatbolGamma$, the envelope estimator of $\boltheta$  is set as $\hatboltheta_{\Env}=\hatbolGamma\hatbolGamma\T\hatboltheta=\mbP_{\hatbolGamma}\hatboltheta$. Given the true envelope dimension $u$ and the $\sqrt{n}$-consistent standard or initial estimators $\hatmbM$ and $\hatmbU=\hatboltheta\hatboltheta\T$, $\mbP_{\hatbolGamma}=\hatbolGamma\hatbolGamma\T$ from optimizing the above objective function is a $\sqrt{n}$-consistent estimate for  the projection onto the envelope. Therefore, the envelope estimator $\hatboltheta_{\Env}=\hatbolGamma\hatbolGamma\T\hatboltheta$ is $\sqrt{n}$-consistent and can be much accurate than the standard estimator $\hatboltheta$. In most applications, $\sqrt{n}$-consistent $\hatmbM$ and $\hatmbU$ are easy to obtain, but there lacks a theoretically justified method to choose the crucial hyperparameter $u$ under the generality of the envelope methods.

Different choices of $\hatmbM$ and $\hatmbM+\hatmbU$ lead to different envelope methods in the literature. Table \ref{tab:summaryEMU} summarizes
some commonly used sample estimators $\{\hatmbM,\hatmbU\}$ for envelope
regression. We use $\mbS_{\mbA}$ to denote the sample covariance
matrix of a random vector $\mbA$ and use $\mbS_{\mbA\mid\mbB}$ to
denote the sample conditional covariance of $\mbA\mid\mbB$. For the partial envelope method, $\mbX=(\mbX_1,\mbX_2)$, where $\mbX_1$ is the important predictor. For the generalized linear model, $\mbS_{\mbX(W)}$ is the weighted sample covariance defined in \citet{CookZhang2015foundation}, where more detailed discussion on the choices of $\hatmbM$ and $\hatmbU$ can be found.

\begin{table}
{\renewcommand{\arraystretch}{1.35}
\begin{center}
\begin{tabular}{c||c|c|c|c}
 Envelopes & Response  &  $\ \ $ Partial $\ \ $ & Predictor & Generalized Linear Model
\tabularnewline 
 \hline
$\hatmbM$ & $\mbS_{\mbY\mid\mbX}$ & $\mbS_{\mbY\mid\mbX}$ & $\mbS_{\mbX\mid\mbY}$ & $\mbS_{\mbX(W)}$ or $\mbS_{\mbX}$
\tabularnewline
$\hatmbM+\hatmbU$ & $\mbS_{\mbY}$ & $\mbS_{\mbY\mid\mbX_{2}}$ & $\mbS_{\mbX}$ & $\hatmbM+\hatbolbeta\mbS_{\mbX(W)}\hatbolbeta\T$\tabularnewline 
\end{tabular}
\end{center}
}
\caption{\label{tab:summaryEMU} Some commonly used sample estimators for envelope
regression: response envelope \citep{Cook2010envelope}, partial envelope \citep{SuCook2011partial} and predictor envelope \citep{Cook2013PLS} for linear models, and envelopes for generalized linear models \citep{CookZhang2015foundation}.}
\end{table}

When the envelope dimension $u$ becomes large, especially when $p$ is not small, the computation based on the full Grassmannian (FG) optimization of \eqref{obj_sample} can be expensive and requires good initial values to circumvent the issue with local minima. When selecting the envelope dimension, this computational issue is even worse: we need to conduct the optimization repeatedly for $k=1,\ldots,p$ since the solutions is not nested as we increase $k$, that is, $\spn(\hatbolGamma_k)\nsubseteq\spn(\hatbolGamma_{k+1})$. Thus, in Section~\ref{subsec:1D} we propose a computationally efficient alternative to the FG envelope dimension selection approach that is based on FG optimization of \eqref{obj_sample}. Our new approach is based on the 1D algorithm proposed by \citet{CookZhang2015algorithm} that breaks down the FG optimization of (\ref{obj_sample}) to ``one-direction-at-a-time''.  We review the population 1D algorithm in the following. 

For $k=0,\dots,p-1$, let $\mbg_{k}\in\mbbR^{p}$ denote the $k$-th sequential direction to be obtained. Let $\mbG_{k}=(\mbg_{1},\dots,\mbg_{k})$, and $(\mbG_{k},\mbG_{0k})$
be an orthogonal basis for $\mbbR^{p}$ and set initial value $\mbg_{0}=\mbG_{00}=0$. Define $\mbM_{k}=\mbG_{0k}\T\mbM\mbG_{0k}$, $\mbU_{k}=\mbG_{0k}\T\mbU\mbG_{0k}$, and the objective function after $k$ sequential steps 
\begin{equation}
\phi_{k}(\mbw)=\log(\mbw\T\mbM_{k}\mbw)+\log\{\mbw\T(\mbM_{k}+\mbU_{k})^{-1}\mbw\},\label{phi_k}
\end{equation}
which has to be minimized over $\mbw\in\mbbR^{p-k}$ subject to $\mbw\T\mbw=1$. The $(k+1)$-th envelope direction is $\mbg_{k+1}=\mbG_{0k}\mbw_{k+1}$, where $\hatmbw_{k+1}=\arg\min_{\mbw\T\mbw=1}\phi_{k}(\mbw)$. The 1D algorithm produces a nested solution path that contains the true envelope: $\spn(\mbG_1)\subset\dots\subset\spn(\mbG_u)=\MenvU\subset\spn(\mbG_{u+1})\subset\dots\subset\spn(\mbG_p)=\mbbR^p$.
As we replace $\mbM$ and $\mbU$ in the above optimization with some $\sqrt{n}$-consistent $\hatmbM$ and $\hatmbU$, we will obtain sequential $\sqrt{n}$-consistent estimates $\hatmbG_{k}=(\hatmbg_{1},\dots,\hatmbg_{k})\in\mbbR^{p\times k}$, $k=1,\dots,p$.

\section{Envelope Dimension Selection}

\subsection{A new quasi-likelihood argument for model-free envelope estimation}\label{subsec:quasi}

The generic moment-based envelope estimation of $\boltheta$ is essentially a two-stage estimator, where the first stage is estimating an envelope basis $\hatbolGamma$ from $\Jn(\bolGamma)$ and the second stage is projecting the standard estimator onto the estimated envelope subspace: $\hatboltheta_{\Env}=\hatbolGamma\hatbolGamma\T\hatboltheta$ to eliminate immaterial variation.
The objective function $\Jn(\bolGamma)$ has previously been proposed and studied by \citet{CookZhang2015algorithm} and \citet{CookZhang2015foundation} purely for estimating an envelope basis, but it is still difficult to understand the effect and implication of $\Jn(\bolGamma)$ on $\hatboltheta_{\Env}$ and to study the asymptotic distribution of $\hatboltheta_{\Env}$. 

We show that $\Jn(\bolGamma)$ can be viewed as a quasi-likelihood function. Moreover, our results connect $\Jn(\bolGamma)$ with the joint estimation of $\mbM$ and $\boltheta$ that leads to both the standard and the envelope estimators. Define
\begin{equation}
\Ln(\mbM,\boltheta)=\log\vert\mbM\vert+\tr\left[\mbM^{-1}\left\{ \hatmbM+(\hatboltheta-\boltheta)(\hatboltheta-\boltheta)\T\right\} \right].\label{Ln}
\end{equation}
Then, given a working dimension $k=0,\dots,p,$ that is not necessarily the true envelope dimension $u$, the envelope estimation is a constrained minimization of \eqref{Ln} over the following feasible parameter set,
\begin{equation}\label{Au}
\mathcal{A}_k=\{(\mbM,\boltheta): \mbM=\bolGamma\bolOmega\bolGamma\T+\bolGamma_{0}\bolOmega_{0}\bolGamma_{0}\T>0,\  \boltheta=\bolGamma\boleta,\boleta\in\mathrm{R}^{k\times 1},\ \mathrm{and}\ (\bolGamma,\bolGamma_0)\T(\bolGamma,\bolGamma_0)=\mbI_p
\},
\end{equation}
where $\mathcal{A}_0$ is defined as $\mathcal{A}_0=\{(\mbM,\boltheta): \mbM>0,\boltheta=0\}$, and the standard estimator is achieved at $\mathcal{A}_p$.

Under the envelope parametrization of $\mbM=\mbM(\bolGamma,\bolOmega,\bolOmega_{0})$ and $\boltheta=\boltheta(\bolGamma,\boleta)$ in \eqref{Au}, $\Ln(\mbM,\boltheta)$ in \eqref{Ln} is now an over-parametrized objective function for the envelope estimation: $\Ln(\mbM,\boltheta)=\Ln(\bolGamma,\bolOmega,\bolOmega_0,\boleta)$. We show that this constrained optimization problem reproduces $\Jn(\bolGamma)$ and $\hatboltheta_{\mathrm{Env}}$  in \citet{CookZhang2015foundation}. 
\begin{lem} \label{lem: Ln} The minimizer of $\Ln(\mbM,\boltheta)$
in \eqref{Ln} under the envelope parametrization in  \eqref{Au} is $\widehat{\mbM}_{\Env}={\hatbolGamma}\hatbolGamma\T\hatmbM{\hatbolGamma}\hatbolGamma\T+{\hatbolGamma_0}\hatbolGamma_0\T\hatmbM{\hatbolGamma_0}\hatbolGamma_0\T$
and $\hatboltheta_{\Env}={\hatbolGamma}\hatbolGamma\T\hatboltheta$, where
$\hatbolGamma$ is the minimizer of the partially optimized objective
function $\Ln(\bolGamma)=\min_{\bolOmega,\bolOmega_0,\boleta}\Ln(\bolGamma,\bolOmega,\bolOmega_0,\boleta)=\Jn(\bolGamma)+\log\vert\hatmbM+\hatmbU\vert+p$ for $\hatmbU=\hatboltheta\hatboltheta\T$.
\end{lem} 
Lemma~\ref{lem: Ln} shows that, although $\Jn(\bolGamma)$ is not an objective function for $\boltheta$, it can be viewed as a partially minimized quasi-likelihood function $\Ln(\mbM,\boltheta)$ under the envelope parametrization, up to an additive constant difference. Our dimension selection method is based on this quasi-likelihood formulation that is completely generic and model-free. This new finding and formulation will largely facilitate our theoretical derivation of envelope dimension selection consistency in the next two sections.
%For ease of notation, we will still use $\Jn(\bolGamma)$ instead of $\Ln(\bolGamma)$.

\subsection{Dimension selection based on the Full Grassmannian optimization}

We first discuss some properties about $\Jn(\bolGamma)$ defined in \eqref{obj_sample} to motivate our dimension selection criterion. It can be shown that $\Jn(\bolGamma)$ converges uniformly in probability to its population counterpart $\J(\bolGamma)=\log\vert\bolGamma\T\mbM\bolGamma\vert+\log\vert\bolGamma\T(\mbM+\mbU)^{-1}\bolGamma\vert$. To distinguish estimators at different envelope working dimensions,  
 let $\bolGamma_{k}$ and $\hatbolGamma_{k}\in\mbbR^{p\times k}$ denote the minimizers of the population objective function $\J(\bolGamma)$ and the sample objective function $\Jn(\bolGamma)$ at dimension $k$. 
%We have the following lemma summarizes some properties of the objective function.
The objective functions $\Jn(\bolGamma)$ and $\J(\bolGamma)$ are well-defined only for envelope dimension $k=1,\dots,p$. But \eqref{Ln} and \eqref{Au} are well-defined for $k=0$. For $k=0$, we can show that $\min_{\mathcal{A}_0}\Ln(\mbM,\boltheta)=\log\vert\hatmbM+\hatmbU\vert+p$ is achieved at $\hatmbM_{\Env,0}=\hatmbM$ and $\hatboltheta_{\Env,0}=0$. Therefore, we define $\Jn(\bolGamma_k)=\J(\bolGamma_k)=0$ for $k=0$. Consequently, we have the following results.
\begin{lem} \label{lem: J_Gamma_k} If $u=0$, then $\J(\bolGamma_k)=0$ for all $k=0,\dots,p$. If $u>0$, then $\J(\bolGamma_{u})<\J(\bolGamma_{k})<0$, for $0<k<u$, and $\J(\bolGamma_{k})=\J(\bolGamma_{u})<0$, for $k\geq u$.  Moreover, for $0\leq u<k$, $\calE_{\mbM}(\mbU)\subset\spn(\bolGamma_{k})$.
\end{lem}
Lemma~\ref{lem: J_Gamma_k} shows that, $\J(\bolGamma_{k})$ is strictly greater than $\J(\bolGamma_{u})$ when $k<u$, and remains constant once $k$ exceeds $u$. We thus propose to select the envelope dimension
via minimizing the following criterion, 
\begin{equation}
\I(k)\equiv\Jn(\hatbolGamma_{k})+\frac{C\cdot k\cdot\log(n)}{n},\quad k=0,1,\dots,p,\label{BIC_FG}
\end{equation}
where $C>0$ is a constant and $\I(0)=0$. We will discuss more about the choice of $C$ later in Section~\ref{subsec:roleC}. The envelope dimension is selected as $\widehat{u}_{\mathrm{FG}}=\arg\min_{0\leq k\leq p}\I(k)$, where we use subscript $\mathrm{FG}$ to denote full Grassmannian optimization of $\Jn(\bolGamma)$. The criterion \eqref{BIC_FG} has a form similar to the Bayesian information criterion, but has the fundamental difference that $J_n(\hatbolGamma_k)$ is not a likelihood function. Properties of $\I$ are not easy to obtain, as the results for likelihood functions do not apply here. Nevertheless, we can show that \eqref{BIC_FG} leads to consistent dimension selection without likelihood arguments.

\begin{thm} \label{thm: consistency_FG} For any constant $C>0$ and $\sqrt{n}$-consistent $\hatmbM$ and $\hatmbU$ in \eqref{BIC_FG}, we have $\Prob(\widehat{u}_{\mathrm{FG}}=u)\rightarrow1$
as $n\rightarrow\infty$. \end{thm}

We have three remarks about the results in Theorem~\ref{thm: consistency_FG}. First, Theorem~\ref{thm: consistency_FG} reveals that the choice of $C$ does not affect the consistency of our proposed dimension selection procedure. We will discuss more on the role of this constant $C$ in Section~\ref{subsec:roleC}. Second, the consistency shown in Theorem~\ref{thm: consistency_FG} does not require any model assumptions. Therefore, \eqref{BIC_FG} can be applied to any models with the envelope structure. Third, in the heavily-studied case of multivariate linear regression model, $\Jn(\bolGamma)$ will reproduce the normal likelihood-based objective function if we plug in appropriate choices of $\hatmbM$ and $\hatmbU$ \citep[Section 1.3;][]{CookZhang2015foundation}. In such cases, \eqref{BIC_FG} will reproduce the Bayesian information criteria for multivariate linear envelope models, where the same criterion in \eqref{BIC_FG} has been used without any justification but yielded good results. The following Corollary to Theorem~\ref{thm: consistency_FG} confirmed that the envelope dimension $\widehat{u}_{\mathrm{BIC}}$ selected from the Bayesian Information Criterion is indeed consistent. 
\begin{cy} \label{cy: BIC_consistency} Suppose that the sample covariance matrices $\mbS_{\mbX}$, $\mbS_{\mbY}$ and $\mbS_{\mbX\mbY}$ are $\sqrt{n}$-consistent, then for envelope linear models, we have $\Prob(\widehat{u}_{\mathrm{BIC}}=u)\rightarrow1$
as $n\rightarrow\infty$.
\end{cy}
Corollary~\ref{cy: BIC_consistency} reinforces the message that, although envelope estimates are typically constructed under some normality assumptions, normality is generally not essential for the application of envelope estimates. 
%For example, \citet{Cook2010envelope,SuCook2011partial,CFZ2015RRE}
%constructed the envelope estimates under the assumption that \eqref{mlm} is true and the error term is normal, while \citet{CookZhang2015simultaneous,Cook2013PLS} considered the case that $(\mbX,\mbY)$ are jointly normal.
%These 
Previous studies of envelope linear models \citep{Cook2010envelope,SuCook2011partial,Cook2013PLS,CFZ2015RRE} have shown that the envelope estimators obtained
by maximizing the normality-based likelihood function are still $\sqrt{n}$-consistent and asymptotically normal even when the normality assumption is violated and the likelihood is mis-specified. Corollary~\ref{cy: BIC_consistency} further showed that, even when the likelihood function is mis-specified due to non-normality, it can still help with selecting the dimension correctly. To the best of our knowledge, this is the first time in the literature that an envelope dimension selection criterion is justified without stringent likelihood assumptions.
For the same reason, the modified BIC in \citet{li2016tensor} is also able to select the tensor envelope dimension consistently since it's also a special case of our FG criterion. 

\subsection{Dimension selection based on the 1D estimation}\label{subsec:1D}

As mentioned earlier in Section~\ref{sec:review}, the FG optimization can not guarantee nested envelope subspace, $\spn(\hatbolGamma_k)\nsubseteq\spn(\hatbolGamma_{k+1})$, while the 1D algorithm always produces a strictly nested solution path: $\spn(\hatmbG_k)\subset\spn(\hatmbG_{k+1})$.
Therefore, it is an intuitive practice \citep[e.g.]{CookZhang2015simultaneous, li2016tensor} to select envelope dimension based on BIC using the 1D envelope estimator.
However, simply replacing $\hatbolGamma_k$ with the 1D estimator $\hatmbG_k$ in BIC, or the FG criterion in general \eqref{BIC_FG}, may not produce asymptotically consistent envelope dimension selection results since $\hatmbG_k$ is not a local optimizer of $\J_n(\bolGamma)$. Therefore, when applying the 1D algorithm, we propose to select the envelope dimension via minimizing the following 1D criterion instead of the FG criterion, 
\begin{equation}
\II(k)\equiv\sum_{j=1}^{k}\phi_{j,n}(\hatmbw_{j})+\frac{C \cdot k\cdot\log(n)}{n},\quad k=0,1,\dots,p,\label{BIC_1d}
\end{equation}
where $C>0$ is a constant, $\II(0)=0$, and the function $\phi_{j,n}(\mbw)$ is the sample version of $\phi_{j}(\mbw)$
defined in \eqref{phi_k}. We select the envelope dimension selected as $\widehat{u}_{\mathrm{1D}}=\arg\min_{0\leq k\leq p}\II(k)$. 
\begin{thm} \label{thm: consistency_1D} For any constant $C>0$ and $\sqrt{n}$-consistent $\hatmbM$ and $\hatmbU$ in \eqref{BIC_1d}, $\Prob(\widehat{u}_{\mathrm{1D}}=u)\rightarrow1$
as $n\rightarrow\infty$. \end{thm}
We have two remarks about the 1D criterion $\II(k)$. First, it is easy to see that $\sum_{j=1}^{k}\phi_{j,n}(\hatmbw_{j})$ serves as the same role as $\Jn(\hatbolGamma_{k})$ in the full Grassmannian optimization criterion $\I(k)$ in \eqref{BIC_FG}. But the change of criterion here is critical as we have a different optimization problem. In fact, simply replacing $\hatbolGamma_{k}$ in the FG criterion \eqref{BIC_FG} with the 1D solution $\hatmbG_{k}$ will not guarantee consistency in the selection. This is due to the fact that $\hatmbG_{k}$, although  a $\sqrt{n}$-consistent envelope basis estimator, is not a local minima of the full Grassmannian objective function $\Jn(\bolGamma)$. Instead, using $\sum_{j=1}^{k}\phi_{j,n}(\hatmbw_{j})$
is indeed necessary for envelope dimension selection based on the 1D algorithm. Secondly, the computational cost of obtaining $\II(k)$, $k=1,\dots,p,$ is much less than that of $\I(k)$, $k=1,\dots,p$. This is not only because the 1D algorithm is much faster and more stable than the FG optimization, but also due to the sequential nature of the 1D algorithm. For the 1D algorithm, we only need to run it once to estimate the $(p-1)$-dimensional envelope $\hatmbG_{p-1}$ to obtain all the values of $\II(k)$, $k=1,\dots,p$. For the full Grassmannian
approach, it requires estimation of each envelope basis $\hatbolGamma_{1},\dots,\hatbolGamma_{p-1}$
separately and the computation for $\hatbolGamma_{p-1}$ alone can be more costly than obtaining $\hatmbG_{p-1}$. Therefore, in practice, we would strongly recommend using the 1D approach instead of the FG approach, when $p$ is large. Simulation studies in the next section also show that the 1D approach is much more accurate and effective than the FG approach.

\subsection{Role of $C$}\label{subsec:roleC}

Our proposed model-free criteria \eqref{BIC_FG} and \eqref{BIC_1d} are motivated from the BIC, and as we mentioned earlier in Corollary~\ref{cy: BIC_consistency}, the FG criterion \eqref{BIC_FG} indeed includes the BIC for envelope linear models as a special choice. Specifically, the first term $\J_n(\hatbolGamma_k)$ in \eqref{BIC_FG} will be the $(2/n)$ times the negative normal log-likelihood with appropriate choices of $\hatmbM$ and $\hatmbU$, whereas the second term $Ck\log(n)/n$ corresponds to the penalty term on the number of parameters in the linear envelope models.
In the 1D criterion, the first term has no likelihood interpretation but is analogous to the first term in the FG criterion, thus the same penalty on the number of parameters were used.
Because of these connections and connections with BIC, we suggest to use $C=1$ for both the FG and the 1D criteria when the parameter $\boltheta$ is naturally a vector.
In other situations, where $\boltheta$ is naturally a matrix-valued or even tensor-valued parameters, we would try to match $Ck$ with the number of parameters in the model.

Although we focused on a vector-valued parameter $\boltheta\in\mbbR^p$ in the quasi-likelihood argument of $\Ln(\mbM,\boltheta)$, the theoretical results in Lemma~\ref{lem: Ln} and Theorems~\ref{thm: consistency_FG} \& \ref{thm: consistency_1D} do not impose any restriction on $\boltheta$ being a vector. Our proofs are in fact written for a matrix-valued $\boltheta\in\mbbR^{p\times q}$ and can be straightforwardly extended to tensor-valued $\boltheta$. In such cases, matching $Ck$ with the number of parameters would give $C=q$ for $\boltheta\in\mbbR^{p\times q}$ when we are enveloping the column space of $\boltheta$. 
Also, the term $(\hatboltheta-\boltheta)(\hatboltheta-\boltheta)\T$ in $\Ln(\mbM,\boltheta)$ may also be replaced by a weighted version $(\hatboltheta-\boltheta)\mbW(\hatboltheta-\boltheta)\T$ for some $\mbW\in\mbbR^{q\times q}$ to tie more closely to the likelihood function for potentially improved efficiency. See \citet{CookZhang2015foundation} (Definition 4 and Proposition 7 in the Supplement) for a detailed discussion on enveloping a matrix-valued parameter and choices of $\mbW>0$. 

The proposed envelope dimension selection approaches in this paper are as flexible as possible, since we only require $\sqrt{n}$-consistent $\hatmbM$ and $\hatmbU$ for the envelope $\MenvU$ without additional assumptions on distributions of variables or specific models.
The theoretical developments, i.e. Theorems~\ref{thm: consistency_FG} and \ref{thm: consistency_1D}, only require $C$ to be a positive constant to guarantee asymptotically correct selection of the envelope dimension with probability one.
However, in finite sample, the selection of envelope dimension may be affected by the choice of $C$. 
It is hard to describe qualitatively the effect $C$ on the dimension selection, because that depends on many factors such as the signal-to-noise ratio of the data, the sample size, the total number of parameters, the efficiency and the variance of the $\sqrt{n}$-consistent estimators $\hatmbM$ and $\hatmbU$, etc.
Nonetheless, from the proposed criteria \eqref{BIC_FG} and \eqref{BIC_1d}, we know that smaller $C$ leads to a more conservative choice of the envelope dimension, potentially overestimation ($\widehat{u}> u$), and larger $C$ leads to a more aggressive choice and potentially underestimation ($\widehat{u}< u$). 
From our experience, the number $C$ should be set to its default value $C=1$ when there is no additional model assumption or prior information. When we know additional model assumption or prior information, $C$ should be set such that $Ck$ best matches the degree-of-freedom or total number of free parameters of the model or estimation procedure. For example,if the envelope is enveloping a vector-valued parameter, e.g. linear or generalized linear regression with univariate response or predictor, then let $C=1$;   if the envelope is enveloping a matrix or tensor valued parameter, then usually the best result comes from $C>1$, where $Ck$ should be obtained from calculating the total number of free parameters, which relate to the dimension of the matrix/tensor as well as the true rank of the parameter matrix/tensor \citep[e.g.]{CFZ2015RRE}.
In the next section, we use $C=1$ for generic envelopes, where we have no information about the model, and also use $C=1$ for envelope models (simulation Section~\ref{sim:models} and real data Section~\ref{real}) where the parameter of interest is a vector; then in Section~\ref{sim:matrix} we study the effect of $C$ for a matrix valued parameter. The numerical results further support our opinion in the above.

\section{Numerical studies}

\subsection{Generic Envelopes\label{sec:sim_generic}}

\begin{table}
\begin{centering}
{\renewcommand{\arraystretch}{1.25}
\begin{tabular}{cccc|cccccc}
 & \multicolumn{3}{c|}{1D selection} & \multicolumn{6}{c}{FG selection}\tabularnewline
 & (I) & (II) & (III) & \multicolumn{2}{c}{(I)} & \multicolumn{2}{c}{(II)} & \multicolumn{2}{c}{(III)}\tabularnewline
$n$ & \multicolumn{3}{c|}{$\widehat{u}_{\mathrm{1D}}=5$} & $\widehat{u}_{\mathrm{FG}}=5$ & $\widehat{u}_{\mathrm{FG}}=6$ & $\widehat{u}_{\mathrm{FG}}=5$ & $\widehat{u}_{\mathrm{FG}}=6$ & $\widehat{u}_{\mathrm{FG}}=5$ & $\widehat{u}_{\mathrm{FG}}=6$\tabularnewline
$150$ & 98 & 45 & 67 & 59.5 & 24 & 66 & 12.5 & 28.5 & 44.5\tabularnewline
$200$ & 99 & 75.5 & 94 & 66 & 23 & 77.5 & 15.5 & 39.5 & 47\tabularnewline
$250$ & 100 & 95 & 97 & 70 & 24 & 82.5 & 14.5 & 33 & 48\tabularnewline
$300$ & 100 & 99.5 & 99 & 67.5 & 24 & 85 & 13.5 & 40.5 & 47\tabularnewline
$400$ & 100 & 100 & 100 & 75.5 & 18.5 & 84.5 & 14 & 49 & 41\tabularnewline
$800$ & 100 & 100 & 100 & 84.5 & 13.5 & 91 & 8.5 & 56 & 39.5\tabularnewline
\end{tabular}
}
\par\end{centering}

\protect\caption{\label{tab:generic} Frequencies of selected dimensions for a generic $\protect\calE_{\protect\mbM}(\protect\mbU)$
with $p=20$ and $u=5$. }
\end{table}

In this section, we present numerical studies of dimension
selection for a generic envelope $\calE_{\mbM}(\mbU)=\spn(\bolGamma)$,
where $\mbM=\bolGamma\bolOmega\bolGamma\T+\bolGamma_{0}\bolOmega_{0}\bolGamma_{0}\T$
and $\mbU=\bolGamma\bolPhi\bolGamma\T$ follow the envelope structure.
In this section, we use $p=20$ and $u=5$. The envelope basis matrix
$\bolGamma\in\mbbR^{p\times u}$ is a randomly generated semi-orthogonal
matrix and then $\bolGamma_{0}\in\mbbR^{p\times(p-u)}$ is the orthogonal
completion of $\bolGamma$ such that $(\bolGamma,\bolGamma_{0})$
is orthogonal.  For $\bolPhi$, we generate $\mbA\in\mathbb{R}^{u\times u}$ with each element $a_{ij}$ sampled from the uniform distribution over [0,1]. Then we set $\bolPhi=\mbA\mbA\T$.
We considered the following three different models
for the symmetric positive definite matrices $\bolOmega$ and $\bolOmega_{0}$. Model (I): both $\bolOmega$ and $\bolOmega_{0}$ are randomly
generated independently in the same way as $\bolPhi$. Model (II): $\bolOmega$ and
$\bolOmega_{0}$ are each generated as $\mbO\mbD\mbO\T$ with $\mbO$ being an orthogonal matrix and $\mbD$ being a diagonal matrix of positive elements on its diagonal. We set the diagonal elements in $\mbD$ for
$\bolOmega$ as $1,\dots,u$, and the diagonal elements in $\mbD$ for $\bolOmega_{0}$ as $\exp(-4),\ \exp(-3.5),\ldots,\exp(3)$. Model (III): all parameters are the same as Model (II) except that $\bolOmega_0$ is now $0.1\mbI_{p-u}$.

We simulated 200 pairs of sample matrices from Wishart distributions,
$\hatmbM\sim W_{p}(\mbM/n,\ n)$ and $\hatmbU\sim W_{p}(\mbU/n,\ n)$
so that they are $\sqrt{n}$-consistent for their population counterparts.
We vary the sample size $n$ from 150 to 800. In 
Table~\ref{tab:generic}, we report the percentages of selecting the envelope dimension correctly by the two proposed approaches. In all three models, the 1D criterion is very effective and provides consistent selection of $u$: the percentage of correctly selecting the envelope dimension is monotonically approaching 1 as the sample size increases. The FG criterion is less competitive but still gives reasonable results especially because the total number of free parameters in $\bolGamma$, $\bolOmega$, $\bolOmega_{0}$ and $\bolPhi$ is $p(p+1)/2+u(u+1)/2=225$ which is not a small number comparing to $n$. For the FG approach, we also reported the percentage of $\widehat{u}_{\mathrm{FG}}=6$ in Table~\ref{tab:generic}, which demonstrates a clear tendency for the FG approach to over-estimate the envelope dimension. From Lemma~\ref{lem: J_Gamma_k}, the over-estimated envelope dimension will result in a larger subspace that contains the true envelope. Thus over-estimating $u$ eventually still leads to consistent and unbiased envelope estimator for $\boltheta$ and cause much less harm than under-estimating $u$.

\subsection{Envelope Models}\label{sim:models}

\begin{table}
\begin{centering}
{\renewcommand{\arraystretch}{1.25}
\begin{tabular}{cccccc|ccccc}
 &  & \multicolumn{4}{c|}{Correct Selection $\%$} & \multicolumn{4}{c}{Estimation Error $\Vert\hatbolbeta-\bolbeta\Vert_{F}$} & \tabularnewline
 &  & \multicolumn{2}{c}{} &  &  & Standard & \multicolumn{3}{c}{Envelope} & \tabularnewline
Model & $n$ &  & 1D & FG &  &  & true $u$ & 1D & FG & S.E.$\leq$\tabularnewline
\multirow{3}{*}{Linear} & 150 &  & 93 & 81 &  & 0.49 & 0.31 & 0.33 & 0.33 & 0.015\tabularnewline
 & 300 &  & 99 & 92 &  & 0.32 & 0.19 & 0.19 & 0.20 & 0.008\tabularnewline
 & 600 &  & 99 & 92.5 &  & 0.23 & 0.13 & 0.14 & 0.14 & 0.007\tabularnewline
\multirow{3}{*}{Logistic} & 150 &  & 72 & 77.5 &  & 2.16 & 0.56 & 0.67 & 0.60 & 0.072\tabularnewline
 & 300 &  & 92 & 89.5 &  & 1.40 & 0.34 & 0.35 & 0.34 & 0.042\tabularnewline
 & 600 &  & 98 & 94 &  & 0.98 & 0.22 & 0.22 & 0.24 & 0.030\tabularnewline
\multirow{3}{*}{Cox} & 150 &  & 58 & 54 &  & 1.33 & 1.24 & 1.22 & 1.23 & 0.022\tabularnewline
 & 300 &  & 83 & 75.5 &  & 0.98 & 0.90 & 0.89 & 0.90 & 0.013\tabularnewline
 & 600 &  & 100 & 93 &  & 0.79 & 0.72 & 0.72 & 0.72 & 0.008\tabularnewline
\end{tabular}
}
\par\end{centering}

\protect\caption{\label{tab:models}  Selection and estimation results for three different envelope models. Left panel includes percentages of correct selection. Right panel includes means and standard errors of $\Vert\protect\hatbolbeta-\protect\bolbeta\Vert_{F}$ for the standard estimator and the envelope estimators with either true or estimated dimensions. }
\end{table}

In this section, we simulate three different envelope models where the envelope dimension is $u=2$ for $p=10$. The three models are: the multivariate linear model \eqref{mlm}, the logistic regression model and the Cox proportional hazard model. For the linear regression in \eqref{mlm}, we generated $X_{i}$ and $\epsilon_i$ independently from $N(0,1)$ and $N_{p}(0,\bolSigma)$, where $\bolbeta=\bolGamma\boleta$, $\boleta=(1,1)\T$, and $\bolSigma=\bolGamma\bolOmega\bolGamma\T+\bolGamma_{0}\bolOmega_{0}\bolGamma_{0}\T$. The covariance $\bolOmega$ and $\bolOmega_{0}$ are each generated as $\mbO\mbD\mbO\T$ similar to Model (II) in Section~\ref{sec:sim_generic}. We set the eigenvalues
as $1,5$ in $\bolOmega$, and $\exp(-4),\ \exp(-3),\ldots,\exp(3)$ in $\bolOmega_{0}$. For the logistic regression: $Y_{i}\sim\mathrm{Bernoulli}(\mathrm{logit}(\bolbeta\T\mbX_{i}))$, we simulate $\mbX_{i}$ from $N_{p}(0,\bolSigma_{\mbX})$ where the parameters $\bolSigma_{\mbX}$
and $\bolbeta$ are the same as $\bolSigma$ and $\bolbeta$ in the linear model. For the Cox model, we follow the simulation model in \citet{CookZhang2015foundation} and let the survival
time follow a Weibull distribution with scale parameter $\exp(\bolbeta\T\mbX/5)$
and shape parameter 5, which gives hazard rate $h(Y\mid\mbX)=5Y^{4}\cdot\exp(\bolbeta\T\mbX)$.
The censoring variable $\delta_{i}$ is generated from Bernoulli(0.5) distributions, which gives censoring rates of approximately 50\%. Then the data $(Y_{i},\delta_{i},\mbX_{i})$, $i=1,\dots,n$, are used to fit the envelope Cox model, where $Y_{i}$ is the failure time, $\delta_{i}=0$ or 1 indicating whether the failure
time is censored or observed, $\mbX_{i}$ is the predictor vector. Data generation for $\mbX_{i}$ is similar to the logistic regression set-up, except for $\bolOmega_{0}=0.1\mbI_{8}$ and $\boleta=(0.2,0.2)\T$.
%The choice of $\hatmbM$ and $\hatmbU$ are $\mbS_{\mbX}$ and $\hatbolbeta\hatbolbeta\T$, respectively.

For each of the above models, we consider sample size $n=150$, 300 and 600 and generated 200 data sets for each of the sample sizes. Table~\ref{tab:models} summarizes the percentages of correctly selected envelope dimension and the estimation error $\Vert\hatbolbeta-\bolbeta\Vert_F$ in each of the simulations, where we compare the standard estimators (e.g. least squares, likelihood and partial
likelihood estimators) to the envelope estimators using the true dimension or using the selected dimensions. 
For all scenarios, there is no significant difference among envelope estimators (whether with the true or the estimated dimension), which are all significantly better than the standard estimator. Regarding dimension selection accuracy, both 1D and FG procedures have produced satisfactory results. The percentage of correct selection is low only for the Cox model at sample size $n=150$, which is a small number considering the 50\% censoring rate. 

\subsection{Matrix-valued parameter}\label{sim:matrix}

\begin{table}
\begin{centering}
{\renewcommand{\arraystretch}{1.25}
\begin{tabular}{c|cccccc|cccccc}
 & \multicolumn{6}{c|}{Correct Selection $\%$} & \multicolumn{6}{c}{Average selected $\widehat{u}$} \tabularnewline
 & \multicolumn{3}{c}{1D} & \multicolumn{3}{c|}{FG} & \multicolumn{3}{c}{1D} & \multicolumn{3}{c}{FG}\tabularnewline
$n$ & 150 & 300 & 600 & 150 & 300 & 600 & 150 & 300 & 600 & 150 & 300 & 600\tabularnewline
\hline 
$C=1$ & 38 & 59 & 63 & 23 & 42 & 52 & 2.95 & 2.53 & 2.46 & 3.51 & 2.90 & 2.62\tabularnewline
$C=3$ & 92 & 100 & 100 & 92 & 100 & 100 & 1.94 & 2.00 & 2.00 & 2.00 & 2.00 & 2.00\tabularnewline
$C=5$ & 66 & 100 & 100 & 86 & 100 & 100 & 1.66 & 2.00 & 2.00 & 1.86 & 2.00 & 2.00\tabularnewline
$C=10$ & 5 & 55 & 100 & 19 & 95 & 100 & 1.05 & 1.55 & 2.00 & 1.19 & 1.95 & 2.00\tabularnewline
\end{tabular}
}
\par\end{centering}

\protect\caption{\label{tab:constantC} Multivariate linear regression, response envelope model with multivariate response of dimension $p=10$ and envelope dimension $u=2$. The parameter of interest is the $10\times3$ regression coefficient matrix, where $q=3$ is the number of predictors, and hence the best choice of $C$ should be $C=q=3$. The left panel summarizes percentages of correct selection; and the right panel summarizes the average of selected dimension.  }
\end{table}

As an illustration of the effect of $C$, we simulated data from the multivariate linear regression model \eqref{mlm}, where we considered the response envelope model with multivariate response of dimension $p=10$ and envelope dimension $u=2$. The parameter of interest is the $10\times3$ regression coefficient matrix, where $q=3$ is the number of predictors, and hence from our discussion in Section~\ref{subsec:roleC} we expect the best choice of $C$ to be $C=q=3$. The parameters and data are generated in same way as the single predictor linear model in Section~\ref{sim:models}, where we still set elements in $\boleta\in\mbbR^{u\times q}$ as all ones to get $\bolbeta=\bolGamma\boleta$.
From Table~\ref{tab:constantC}, we have the following observations: (1) for all values of $C$, the percentage of correct selection goes toward 1 when the sample size goes to infinity (even for $C=1$, the percentage goes slowly but steadily to 90\% as we keep increase the sample size to 6000); this also numerically verifies our theoretical results; (2) the ``best'' choice is apparently $C=3$ because this lets the penalty $Ck$ in the criteria matches the number of parameters in the model; (3) from the average value of selected $\widehat{u}$, we see that $C>3$ leads to underestimation of $u$ when the sample size is small and $C<3$ leads to overestimation; (4) $C=1$ with the 1D criterion is a ``robust'' choice, even for the small sample size $n=150$ the averaged selection is $2.95$. 

We make two additional remarks on the performance of $C=1$. On one hand, the average dimension is only slightly larger than the true dimension even for small sample size. In the situations when $C=1$ is not the optimal choice, the 1D criterion with $C=1$ may overestimate the dimension by a small amount. On the other hand, overestimation of the dimension slightly is much less of an issue comparing to the issue of underestimating the envelope dimension. If we apply envelope methods with a slightly larger structural dimension, estimation of the parameter is still unbiased. The slightly larger structural dimension will only lead to some efficiency loss. Meanwhile, if the dimension is underestimated, the envelope estimator will be biased and important directions will be missed. Fortunately, underestimation is not likely to happen, according to the simulation results in Table~\ref{sim:matrix}.

\subsection{Real data illustration}\label{real}

\begin{figure}[ht!]
\begin{centering}
\includegraphics[bb=100bp 230bp 512bp 542bp,scale=0.8]{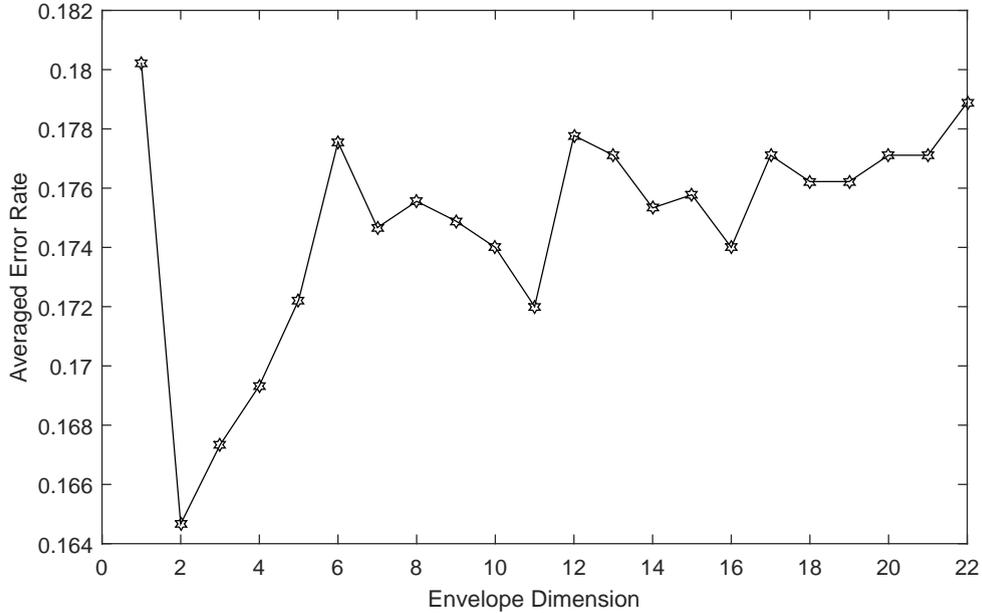}
\par\end{centering}
\protect\caption{\label{fig:realdata} Colon cancer tissue data: averaged mis-classification error rates for moment-based envelope estimators with various dimension based on 100 random data splitting. The standard logistic regression is the rightmost point, where $u=p=22$.}
\end{figure}

For a real data illustration, we revisit the data set for envelope logistic regression in \citet{CookZhang2015foundation}. The data is from a colonoscopy study where 105 adenomatous (precancerous) tissue samples and 180 normal tissue samples were illuminated with ultraviolet light so that they fluoresce at difference wavelengths. The purpose of the study is to classify the total $n=285$ tissue samples into the two classes, i.e. $Y=1$ (adenomatous) and $Y=0$ (normal), using the $p=22$ predictors that are from laser-induced fluorescence spectra measured as 8nm spacing from 375 to 550 nm.  More details of such colonoscopy study and a similar data set can be found in \citet{hawkins2013}.

For this data set, we study the moment-based envelope estimators based on the 1D algorithm as an alternative to the likelihood-based approach demonstrated in \citet{CookZhang2015foundation}. 
Using the model-free dimension selection criterion developed in this paper, envelope dimension $u=2$ is selected by both the FG approach \eqref{BIC_FG} and the 1D approach \eqref{BIC_1d} developed in this paper. 
We then randomly split the data into 80\% training samples (228 samples) and 20\% testing samples (57 samples) repeatedly for 100 times and fit the 1D moment-based envelope estimator for various dimensions and evaluate its classification power on the testing data set.
As a result, the averaged mis-classification error rate is 0.1647 with standard error 0.0051 for the envelope estimator with selected dimension $u=2$, much better than 0.1802 with standard error 0.0047 from fitting with $u=1$ (if we underestimate the envelope dimension), and also much better than 0.1789 with standard error 0.0054 from the standard logistic regression. 
Figure~\ref{fig:realdata} further summarizes the averaged error rate for envelope estimators with various dimensions from $u=1$ to $u=22$.
Clearly, $u=2$ is the desirable envelope dimension for this data set that is selected by our model-free criteria.

On the other hand, if we assume the predictor is normal then the envelope MLE, given the envelope dimension $u$, can be obtained use the iterative algorithm \citep[Algorithm 1]{CookZhang2015foundation}. 
Standard BIC approach for selecting $u$ is then applicable based on the full likelihood of $(Y,\mbX)$. As a result, $u=1$ is selected. However, envelope MLE with $u=1$ will give bad classification result and \citet{CookZhang2015foundation} also used $u=2$ for their envelope MLE, where the dimension is selected based on five-fold cross-validation.

For this data set, the most likely reason for the standard BIC to fail to select a ``reasonable'' envelope dimension is probably due to the non-normality in the predictors.
While cross-validation is computationally more expensive and has no theoretical justification, our proposed 1D and FG selection approaches can relax the normality assumption and select the asymptotically consistent and practically useful envelope dimension.

%\section{Discussion\label{sec:discussion}}
%
%The proposed envelope dimension selection approaches in this paper are as flexible as possible, since we only require $\sqrt{n}$-consistent $\hatmbM$ and $\hatmbU$ for the envelope $\MenvU$. Although we focused on a vector-valued parameter $\boltheta\in\mbbR^p$ in the quasi-likelihood argument of $\Ln(\mbM,\boltheta)$, the theoretical results in Lemma~\ref{lem: Ln} and Theorems~\ref{thm: consistency_FG} \& \ref{thm: consistency_1D} do not impose any restriction on $\boltheta$ being a vector. Our proofs are in fact written for a matrix-valued $\boltheta\in\mbbR^{p\times q}$. In such cases, a natural choice of the constant $C$ in \eqref{BIC_FG} and \eqref{BIC_1d} would be $C=q$. 
%Also, the term $(\hatboltheta-\boltheta)(\hatboltheta-\boltheta)\T$ in $\Ln(\mbM,\boltheta)$ may also be replaced by a weighted version $(\hatboltheta-\boltheta)\mbW(\hatboltheta-\boltheta)\T$ for some $\mbW\in\mbbR^{q\times q}$.  See \citet{CookZhang2015foundation} (Definition 4 and Proposition 7 in the Supplement) for a detailed discussion on enveloping a matrix-valued parameter and choices of $\mbW>0$. 

%For a tensor (multi-dimensional array-valued) parameter $\boltheta\in\mbbR^{p_{1}\times\cdots\times p_{m}}$, our approach is also applicable.
%Comment that 1D is much better than FG, especially $u$ is not small.
%When Also link to \citet{ZouChen2012BIC} for sparsity.
%The criteria $\I(k)$ in \eqref{BIC_FG} and $\II(k)$ in \eqref{BIC_1d} can be easily modified 

\appendix
%dummy comment inserted by tex2lyx to ensure that this paragraph is not empty%dummy comment inserted by tex2lyx to ensure that this paragraph is not empty%dummy comment inserted by tex2lyx to ensure that this paragraph is not empty
\setcounter{equation}{0}
\global\long\def\theequation{A\arabic{equation}}
 
%
%\section*{Acknowledgments}
%
%The authors thank the Associate Editor and referees for their helpful comments that led to significant improvements in this article. Zhang's research was supported in part by NSF grants DMS-1613154 and CCF-1617691. Mai's research was supported in part by NSF grant CCF-1617691.

\section*{Appendix}

\section{Some useful preparation}

%Proof of Lemma \ref{lem: J_Gamma_k} is similar to that of Lemma 6.3 in \citep{Cook2013PLS} and is thus omitted. 
Proof for Corollary 1 is also omitted as it is straightforward from Theorem~\ref{thm: consistency_FG}. The remaining proofs are provided in this Appendix.
We will need to apply the following Proposition~\ref{pn: FG consistency} and Lemma~\ref{lem: appendix CHS}, which are obtained from \citet[Propositions 2, 3, 5 and 6]{CookZhang2015algorithm} and \citet[Lemmas 6.2 and 6.3]{Cook2013PLS}, for our proofs.
\begin{pn} \label{pn: FG consistency} If $k=u$, then $\spn(\bolGamma_{k})=\spn(\mbG_k)=\calE_{\mbM}(\mbU)$;
if, in addition, $\hatmbM$ and $\hatmbU$ are both $\sqrt{n}$-consistent,
then $\hatbolGamma_{k}\hatbolGamma_{k}\T$ and $\hatmbG_k\hatmbG_k\T$ are both $\sqrt{n}$-consistent for the projection onto $\calE_{\mbM}(\mbU)$.
\end{pn}
\begin{lem}\label{lem: appendix CHS} Suppose that $\mbM>0$ is a $p\times p$ symmetric matrix $(\bolGamma,\bolGamma_{0})$ is an orthogonal
basis matrix for $\mbbR^{p}$, then $\log\vert\mbM\vert =  \log\vert\bolGamma_{0}\T\mbM\bolGamma_{0}\vert-\log\vert\bolGamma\T\mbM^{-1}\bolGamma\vert \leq \log\vert\bolGamma_{0}\T\mbM\bolGamma_{0}\vert+\log\vert\bolGamma\T\mbM\bolGamma\vert$, where the second equality holds if and only if $\spn(\bolGamma)$ is a reducing subspace of $\mbM$. 
\end{lem}

\section{Proof for Lemma \ref{lem: Ln}}

\begin{proof} First, we substitute $\mbM=\bolGamma\bolOmega\bolGamma\T+\bolGamma_{0}\bolOmega_{0}\bolGamma_{0}\T$
and $\boltheta=\bolGamma\boleta$ into $\Ln(\mbM,\boltheta)$ and
expand it explicitly as

\begin{eqnarray*}
\Ln(\mbM,\boltheta) & = & \log\vert\bolOmega\vert+\log\vert\bolOmega_{0}\vert\\
 & + & \tr\left[(\bolGamma\bolOmega^{-1}\bolGamma\T+\bolGamma_{0}\bolOmega_{0}^{-1}\cdot\bolGamma_{0}\T)\cdot\left\{ \hatmbM+(\hatboltheta-\bolGamma\boleta)(\hatboltheta-\bolGamma\boleta)\T\right\} \right]\\
 & \equiv & \Ln(\bolGamma,\bolOmega,\bolOmega_{0},\boleta)
\end{eqnarray*}
where the first part is from $\log\vert\mbM\vert=\log\vert\bolGamma\bolOmega\bolGamma\T+\bolGamma_{0}\bolOmega_{0}\bolGamma_{0}\T\vert=\log\vert\bolOmega\vert+\log\vert\bolOmega_{0}\vert$.
We next show that $\Jn(\bolGamma)$ is obtained by partially minimizing
$\Ln(\bolGamma,\bolOmega,\bolOmega_{0},\boleta)$ over $\boleta$,
$\bolOmega$ and $\bolOmega_{0}$. Taking derivative of $\Ln(\bolGamma,\bolOmega,\bolOmega_{0},\boleta)$
with respect to $\boleta$ and set it equaling zero, we have 
\[
0=(\bolGamma\bolOmega^{-1}\bolGamma\T+\bolGamma_{0}\bolOmega_{0}^{-1}\cdot\bolGamma_{0}\T)\cdot(2\boleta-2\bolGamma\T\hatboltheta),
\]
which leads to the minimizer $\hatboleta(\bolGamma)=\bolGamma\T\hatboltheta$.
As a result, $\hatboltheta(\bolGamma)=\bolGamma\bolGamma\T\hatboltheta=\mbP_{\bolGamma}\hatboltheta$.
Furthermore, the partially minimized $\Ln$ is now
\begin{eqnarray*}
\Ln(\bolGamma,\bolOmega,\bolOmega_{0}) & = & \log\vert\bolOmega\vert+\log\vert\bolOmega_{0}\vert\\
 & + & \tr\left[(\bolGamma\bolOmega^{-1}\bolGamma\T+\bolGamma_{0}\bolOmega_{0}^{-1}\cdot\bolGamma_{0}\T)\cdot\left\{ \hatmbM+(\hatboltheta-\mbP_{\bolGamma}\hatboltheta)(\hatboltheta-\mbP_{\bolGamma}\hatboltheta)\T\right\} \right]\\
 & = & \log\vert\bolOmega\vert+\log\vert\bolOmega_{0}\vert+\tr\left[(\bolGamma\bolOmega^{-1}\bolGamma\T+\bolGamma_{0}\bolOmega_{0}^{-1}\cdot\bolGamma_{0}\T)\cdot\left\{ \hatmbM+\mbQ_{\bolGamma}\hatboltheta\hatboltheta\T\mbQ_{\bolGamma}\right\} \right]\\
 & = & \log\vert\bolOmega\vert+\log\vert\bolOmega_{0}\vert+\tr(\bolGamma\bolOmega^{-1}\bolGamma\T\cdot\hatmbM)\\
 & + & \tr\left\{ \bolGamma_{0}\bolOmega_{0}^{-1}\bolGamma_{0}\T\cdot(\hatmbM+\mbQ_{\bolGamma}\hatboltheta\hatboltheta\T\mbQ_{\bolGamma})\right\} \\
 & = & \log\vert\bolOmega\vert+\log\vert\bolOmega_{0}\vert+\tr(\bolOmega^{-1}\cdot\bolGamma\T\hatmbM\bolGamma)\\
 & + & \tr\left\{ \bolOmega_{0}^{-1}\cdot\bolGamma_{0}\T(\hatmbM+\mbQ_{\bolGamma}\hatboltheta\hatboltheta\T\mbQ_{\bolGamma})\bolGamma_{0}\right\} .
\end{eqnarray*}
It is a well-known fact (from normal likelihood) that $\mbS=\arg\min_{\bolSigma>0}\{\tr(\bolSigma^{-1}\mbS)+\log\vert\mbS\vert\}$.
This leads to the minimizers $\hatbolOmega(\bolGamma)=\bolGamma\T\hatmbM\bolGamma$
and $\hatbolOmega_{0}(\bolGamma)=\bolGamma_{0}\T(\hatmbM+\mbQ_{\bolGamma}\hatboltheta\hatboltheta\T\mbQ_{\bolGamma})\bolGamma_{0}=\bolGamma_{0}\T(\hatmbM+\hatboltheta\hatboltheta\T)\bolGamma_{0}=\bolGamma_{0}\T(\hatmbM+\hatmbU)\bolGamma_{0}$
from the last equality of $\Ln(\bolGamma,\bolOmega,\bolOmega_{0})$
above. The partially minimized objective function of $\bolGamma$
is finally
\begin{eqnarray*}
\Ln(\bolGamma) & = & \log\vert\hatbolOmega(\bolGamma)\vert+\log\vert\hatbolOmega_{0}(\bolGamma)\vert+u+p-u\\
 & = & \log\vert\bolGamma\T\hatmbM\bolGamma\vert+\log\vert\bolGamma_{0}\T(\hatmbM+\hatmbU)\bolGamma_{0}\vert+p\\
 & = & \log\vert\bolGamma\T\hatmbM\bolGamma\vert+\log\vert\bolGamma\T(\hatmbM+\hatmbU)^{-1}\bolGamma\vert+\log\vert\hatmbM+\hatmbU\vert+p,
\end{eqnarray*}
where the last equality is obtained from Lemma \ref{lem: appendix CHS}.
Thus, we have proven that $\Ln(\bolGamma)=\Jn(\bolGamma)+\log\vert\hatmbM+\hatmbU\vert+p$
and the minimizer $\hatbolGamma$ for $\Jn(\bolGamma)$ is also the
minimizer of the partially minimized negative quasi-likelihood function
$\Ln(\bolGamma)$. It is then straightforward to see that $\hatmbM_{\Env}=\hatbolGamma\hatbolOmega(\hatbolGamma)\hatbolGamma\T+\hatbolGamma_{0}\hatbolOmega_{0}(\hatbolGamma)\hatbolGamma_{0}\T=\mbP_{\hatbolGamma}\hatmbM\mbP_{\hatbolGamma}+\mbQ_{\hatbolGamma}\hatmbM\mbQ_{\hatbolGamma}$
and $\hatboltheta_{\Env}=\hatbolGamma\hatboleta(\hatbolGamma)=\mbP_{\hatbolGamma}\hatboltheta$.

\end{proof}

\section{Proof for Lemma \ref{lem: J_Gamma_k}}

\begin{proof} The Lemma's proof is similar to the proof of Lemma
6.3 of Cook et al. (2013). For completeness, we provide a complete
proof here. For $u=0$, it is clear that $\mbU=0$ and thus $\spn(\bolGamma_{k})$
will be any $k$-dimensional reducing subspace of $\mbM$ for all
$k$ and $\J(\bolGamma_{k})=0=\J(\bolGamma_{u})$. For $u\geq1$,
we write $\J(\bolGamma)$ as
\begin{eqnarray*}
\J(\bolGamma) & = & \log|\bolGamma\T\mbM\bolGamma|+\log|\bolGamma\T(\mbM+\mbU)^{-1}\bolGamma|\\
 & = & \log|\bolGamma\T\mbM\bolGamma|+\log|\bolGamma_{0}\T(\mbM+\mbU)\bolGamma_{0}|-\log\vert\mbM+\mbU\vert\\
 & \geq & \log|\bolGamma\T\mbM\bolGamma|+\log|\bolGamma_{0}\T\mbM\bolGamma_{0}|-\log\vert\mbM+\mbU\vert\\
 & \geq & \log|\mbM|-\log\vert\mbM+\mbU\vert,
\end{eqnarray*}
where the first inequality attains its equality if and only if $\bolGamma_{0}\T\mbU\bolGamma_{0}=0$,
which is equivalent to $\spn(\mbU)\subseteq\spn(\bolGamma)$; the
second inequality attains its equality if and only if $\spn(\bolGamma)$
is a reducing subspace of $\mbM$. Since the envelope $\calE_{\mbM}(\mbU)$
is the smallest subspace satisfying both conditions, $k=u$ is the
minimum dimension for $\J(\bolGamma_{k})$ to achive the minimum $\J(\bolGamma_{u})=\log\vert\mbM\vert-\log\vert\mbM+\mbU\vert<0$.
Hence, $\J(\bolGamma_{u})<\J(\bolGamma_{k})<0$, for $0<k<u$. So
far, we only left to show that the minimum value $\J(\bolGamma_{u})$
is achievable by $\J(\bolGamma_{k})$ for $k>u$. Consider decomposing
$\mbM$ as $\mbM=\bolGamma_{u}\bolOmega\bolGamma_{u}\T+\bolGamma_{0u}\bolOmega_{0}\bolGamma_{0u}\T$,
let $\mbB_{k-u}\in\mbbR^{(p-u)\times(k-u)}$ be a semi-orthogonal
basis for a reducing subspace of $\bolOmega_{0}$. Then by letting
$\bolGamma_{k}$ equal to $(\bolGamma_{u},\mbA_{k-u})$, where $\mbA_{k-u}=\bolGamma_{0u}\mbB_{k-u}\in\mbbR^{p\times(k-u)}$,
it is straightforward to see that $\bolGamma_{k}$ is a reducing subspace
of $\mbM$ that contains $\spn(\mbU)$ thus the minimum of the objective
function is achieved: $\J(\bolGamma_{k})=\J(\bolGamma_{u})$.

\end{proof}

\section{Proof for Theorem \ref{thm: consistency_FG}}

\begin{proof} 

We need to show that $\Prob\left(\I(k)-\I(u)>0\right)\rightarrow1$
as $n\rightarrow\infty$ for both $0\leq k<u$ and $0\leq u<k$ scenarios.
By definition of $\I(k)$, we have 
\begin{equation}
\I(k)-\I(u)=\Jn(\hatbolGamma_{k})-\Jn(\hatbolGamma_{u})+(k-u)\cdot\log(n)/n.\label{Ink_Inu}
\end{equation}

Firstly, for $0\leq k<u$, suffice it to show that $\Jn(\hatbolGamma_{k})-\Jn(\hatbolGamma_{u})=\J(\bolGamma_{k})-\J(\bolGamma_{u})+o_{p}(1)$,
where $\J(\bolGamma_{u})<\J(\bolGamma_{k})<0$ from Lemma \ref{lem: J_Gamma_k}.
We have $\Jn(\hatbolGamma_{j})=\J(\bolGamma_{j})+o_{p}(1)$ for all
$j=1,\dots p,$ because both the sample and population objective functions
are essentially optimized over Grassmannian, i.e. $\bolGamma$ affects
the objective functions $\Jn(\bolGamma)$ and $\J(\bolGamma)$ only
through $\spn(\bolGamma)$. The functions are differentiable and the
derivative $\nabla_{k}\Jn(\hatbolGamma_{k})=\nabla_{u}\Jn(\hatbolGamma_{k})=0$,
where $\nabla_{k}$ and $\nabla_{u}$ are derivatives over the Grassmannians
$Gr(p,k)$ and $Gr(p,u)$, respectively.

Next, for $0\leq u<k$, we show in the following that $\Jn(\hatbolGamma_{k})-\Jn(\hatbolGamma_{u})=\J(\bolGamma_{k})-\J(\bolGamma_{u})+O_{p}(n^{-1})=O_{p}(n^{-1})$.
It follows from \eqref{Ink_Inu} that the dominant term in $\I(k)-\I(u)$
is $(k-u)\cdot\log(n)/n$, which is a positive number. Therefore,
$\Prob\left(\I(k)-\I(u)>0\right)\rightarrow1$ as $n\rightarrow\infty$
for $0\leq u<k$. The special case of $u=0$ is included in the derivation,
as $\J(\bolGamma_{u})=\J(\bolGamma_{k})=0$ for all $k$ and $\bolGamma_{k}$
can be any $k$-dimensional reducing subspace of $\mbM$. 

To show that $\Jn(\hatbolGamma_{k})-\Jn(\hatbolGamma_{u})=O_{p}(n^{-1})$
for $k>u$, we use the negative quasi-likelihood function $\Ln(\mbM,\boltheta)$
in \eqref{Ln}. By Lemma \ref{lem: Ln}, we know that $\Jn(\hatbolGamma_{k})-\Jn(\hatbolGamma_{u})=\Ln(\hatbolGamma_{k})-\Ln(\hatbolGamma_{u})=\Ln(\hatmbM_{\Env,k},\hatboltheta_{\Env,k})-\Ln(\hatmbM_{\Env,u},\hatboltheta_{\Env,u})$,
where $\hatmbM_{\Env}$ and $\hatboltheta_{\Env}$ is defined in Lemma
\ref{lem: Ln} and we use additional subscript $k$ and $u$ to distinguish
different envelope basis $\hatbolGamma$ in $\hatmbM_{\Env}$ and
$\hatboltheta_{\Env}$. We further use $\bolpsi=\{\vech\T(\mbM),\vecc\T(\boltheta)\}\T\in\mbbR^{p(p+1)/2+pq}$
to denote the vector of all unique parameters in the quasi-likelihood
function and write $\Ln(\bolpsi)\equiv\Ln(\mbM,\boltheta)$, and define
$\hatbolpsi$, $\hatbolpsi_{k}$ and $\hatbolpsi_{u}$ from the estimators
$(\hatmbM,\hatboltheta)$, $(\hatmbM_{\Env,k},\hatboltheta_{\Env,k})$
and $(\hatmbM_{\Env,u},\hatboltheta_{\Env,u})$, respectively. To
show $\Ln(\hatbolpsi_{k})-\Ln(\hatbolpsi_{u})=O_{p}(n^{-1})$, we
consider Taylor expansion of $\Ln(\hatbolpsi_{k})$ at $\hatbolpsi_{u}$:
$\Ln(\hatbolpsi_{k})=\Ln(\hatbolpsi_{u})+\Ln^{\prime}(\hatbolpsi_{u})(\hatbolpsi_{k}-\hatbolpsi_{u})+(1/2)(\hatbolpsi_{k}-\hatbolpsi_{u})\T\Ln^{\prime\prime}(\widetilde{\bolpsi}_{u})(\hatbolpsi_{k}-\hatbolpsi_{u})$
and
\[
\Ln(\hatbolpsi_{k})-\Ln(\hatbolpsi_{u})=\Ln^{\prime}(\hatbolpsi_{u})(\hatbolpsi_{k}-\hatbolpsi_{u})+(1/2)(\hatbolpsi_{k}-\hatbolpsi_{u})\T\Ln^{\prime\prime}(\widetilde{\bolpsi}_{u})(\hatbolpsi_{k}-\hatbolpsi_{u}),
\]
where $\widetilde{\bolpsi}_{u}$ is in the neighborhood of $\hatbolpsi_{u}$
so that we can find a series of $\widetilde{\bolpsi}_{u}$ such that
$\Ln^{\prime\prime}(\widetilde{\bolpsi}_{u})$ converge in probability
to a positive definite matrix in probability as $n\rightarrow\infty$.
Since $k>u$, the estimators $\hatbolpsi_{k}$ is unbiased and $\sqrt{n}$-consistent.
Recall that the objective function $\ell_{n}(\mbM,\boltheta)=\log\vert\mbM\vert+\tr[\mbM^{-1}\{\hatmbM+(\hatboltheta-\boltheta)(\hatboltheta-\boltheta)\T\}]$
is smooth and arbitrarily order differentiable with respect to $\mbM>0$
and $\boltheta$, and thus with respect to their unique elements vector
$\bolpsi$. Therefore $\Ln^{\prime}(\bolpsi)$ is a smooth differentiable
function of $\bolpsi$ such that $\Ln^{\prime}(\hatbolpsi_{u})=\Ln^{\prime}(\hatbolpsi)+O_{p}(n^{-1/2})=0+O_{p}(n^{-1/2})$.
For some $\widetilde{\bolpsi}_{u}$ in the neighborhood of $\hatbolpsi_{u}$
that $\widetilde{\bolpsi}_{u}\rightarrow\bolpsi_{u}$ in probability,
$\Ln^{\prime\prime}(\widetilde{\bolpsi}_{u})=O_{p}(1)$. Since both
$\hatbolpsi_{k}$ and $\hatbolpsi_{u}$ is $\sqrt{n}$-consistent
and can be writen as $\bolpsi+O_{p}(n^{-1/2})$, we have $\Ln(\hatbolpsi_{k})-\Ln(\hatbolpsi_{u})=O_{p}(n^{-1/2})*O_{p}(n^{-1/2})+O_{p}(n^{-1/2})*O_{p}(1)*O_{p}(n^{-1/2})=O_{p}(n^{-1})$. 

\end{proof}

\section{Proof for Theorem \ref{thm: consistency_1D}}

\begin{proof} 

We re-write $\II(k)$, $k=1,\dots,p$, as $\II(k)=\sum_{j=1}^{k}\left\{ \phi_{j}(\hatmbw_{j})+\log(n)/n\right\} .$
The increment $\II(k)-\II(k-1)=\phi_{k,n}(\hatmbw_{k})+\log(n)/n$
is exactly the full Grassmannian criterion for the envelope $\calE_{\mbM_{k}}(\mbU_{k})$
evaluated at the one-dimensional envelope estimator. From the following
proof, we will show that the negative term $\phi_{k,n}(\hatmbw_{k})$
dominates the positive term $\log(n)/n$ for $k<u$ because the envelope
$\calE_{\mbM_{k}}(\mbU_{k})$ has dimension greater than 0, then the
positive term $\log(n)/n$ will dominate the negative term for $k>u$
because the envelope $\calE_{\mbM_{k}}(\mbU_{k})$ has dimension zero.
More specifically, we claim that the following two statements are
true: 
\begin{enumerate}
\item for $j\leq u$, $\phi_{j,n}(\hatmbw_{j})+\log(n)/n$ converges to
a negative constant $\phi_{j}(\mbw_{j})<0$, in probability, as $n\rightarrow\infty$;
and 
\item for $j>u$, $\phi_{j,n}(\hatmbw_{j})=O_{p}(n^{-1})$ and $\Prob\left(\phi_{j}(\hatmbw_{j})+\log(n)/n>0\right)\rightarrow1$
as $n\rightarrow\infty$. 
\end{enumerate}
Then the first statement implies that, for $j<u$, $\Prob\left(\II(k)-\II(u)>0\right)\rightarrow1$
as $n\rightarrow\infty$; and the second statement implies that for
$j>u$, $\Prob\left(\II(k)-\II(u)>0\right)\rightarrow1$ as $n\rightarrow\infty$.
The conclusion, $\Prob(\widehat{u}_{\mathrm{1D}}=u)\rightarrow1$
as $n\rightarrow\infty$, thus follows from the above two statements,
which are proved in the following.

From Proposition 4 in \citet{CookZhang2015algorithm}, we know that
$\mbw_{k+1}\in\calE_{\mbM_{k}}(\mbU_{k})$ implies $\mbg_{k+1}=\mbG_{0k}\mbw_{k+1}/||\mbw_{k+1}||\in\calE_{\mbM}(\mbU)$,
and that $\calE_{\mbM_{k}}(\mbU_{k})$ has dimension greater than
zero (i.e. $\calE_{\mbM_{k}}(\mbU_{k})$ not equals to the origin)
if and only if $k\leq u$. Then, for $j\leq u$, the first statement
follows because $\phi_{j,n}(\mbw)$ is a smooth differentiable function
of $\mbw$ and $\hatmbw_{j}$ is $\sqrt{n}$-consistent for $\mbw_{j}$
(in terms of their projection matrices, upon which the functional
value $\phi_{n,j}(\mbw)$ solely depends). The function $\phi_{j,n}(\hatmbw_{j})$
converges to a negative value $\phi_{j}(\mbw_{j})<0$ in probability
as shown in the proof of Propositions 5 and 6 in \citet{CookZhang2015algorithm}.
The proof of Theroem \ref{thm: consistency_FG} only requires  $\hatmbM$
and $\hatmbU$ to be $\sqrt{n}$-consistent estimators. Now $\hatmbM_{k}$
and $\hatmbU_{k}$ are also $\sqrt{n}$-consistent \citep[Proposition 6;][]{CookZhang2015algorithm}.
For $j>u$, the second statement $\phi_{j,n}(\hatmbw_{j})-0=O_{p}(n^{-1})$
can be proved following the lines of proof for Theroem \ref{thm: consistency_FG},
by replacing $\J_{n}(\hatbolGamma)$, $\calE_{\mbM}(\mbU)$ with $\phi_{n,j}(\hatmbw)$
and $\calE_{\mbM_{k}}(\mbU_{k})$ and by noticing this is the special
case of $k=1>u=0$ for Theroem \ref{thm: consistency_FG}.

\end{proof}

\baselineskip=15pt
\bibliography{ref_1Dtest}

\begin{thebibliography}{}

\bibitem[Akaike, 1974]{aic}
Akaike, H. (1974).
\newblock A new look at the statistical model identification.
\newblock {\em IEEE transactions on automatic control}, 19(6):716--723.

\bibitem[Bura and Cook, 2003]{bura2003rank}
Bura, E. and Cook, R.~D. (2003).
\newblock Rank estimation in reduced-rank regression.
\newblock {\em Journal of Multivariate Analysis}, 87(1):159--176.

\bibitem[Conway, 1990]{Conway1990}
Conway, J. (1990).
\newblock {\em A Course in Functional Analysis. Second edition.}
\newblock Springer, New York.

\bibitem[Cook et~al., 2015]{CFZ2015RRE}
Cook, R.~D., Forzani, L., and Zhang, X. (2015).
\newblock Envelopes and reduced-rank regression.
\newblock {\em Biometrika}, 102(2):439--456.

\bibitem[Cook et~al., 2013]{Cook2013PLS}
Cook, R.~D., Helland, I.~S., and Su, Z. (2013).
\newblock Envelopes and partial least squares regression.
\newblock {\em J. R. Stat. Soc. Ser. B. Stat. Methodol.}, 75(5):851--877.

\bibitem[Cook et~al., 2010]{Cook2010envelope}
Cook, R.~D., Li, B., and Chiaromonte, F. (2010).
\newblock Envelope models for parsimonious and efficient multivariate linear
  regression.
\newblock {\em Statist. Sinica}, 20(3):927--960.

\bibitem[Cook et~al., 2004]{cook2004determining}
Cook, R.~D., Li, B., et~al. (2004).
\newblock Determining the dimension of iterative hessian transformation.
\newblock {\em The Annals of Statistics}, 32(6):2501--2531.

\bibitem[Cook and Su, 2013]{cook2013scaled}
Cook, R.~D. and Su, Z. (2013).
\newblock Scaled envelopes: scale-invariant and efficient estimation in
  multivariate linear regression.
\newblock {\em Biometrika}, 100(4):939--954.

\bibitem[Cook and Zhang, 2015a]{CookZhang2015foundation}
Cook, R.~D. and Zhang, X. (2015a).
\newblock Foundations for envelope models and methods.
\newblock {\em Journal of the American Statistical Association},
  110(510):599--611.

\bibitem[Cook and Zhang, 2015b]{CookZhang2015simultaneous}
Cook, R.~D. and Zhang, X. (2015b).
\newblock Simultaneous envelopes for multivariate linear regression.
\newblock {\em Technometrics}, 57(1):11--25.

\bibitem[Cook and Zhang, 2016]{CookZhang2015algorithm}
Cook, R.~D. and Zhang, X. (2016).
\newblock Algorithms for envelope estimation.
\newblock {\em Journal of Computational and Graphical Statistics},
  25(1):284--300.

\bibitem[Eck et~al., 2015]{eck2015application}
Eck, D.~J., Geyer, C.~J., and Cook, R.~D. (2015).
\newblock An application of envelope methodology and aster models.

\bibitem[Geyer et~al., 2007]{geyer2007aster}
Geyer, C.~J., Wagenius, S., and Shaw, R.~G. (2007).
\newblock Aster models for life history analysis.
\newblock {\em Biometrika}, 94(2):415--426.

\bibitem[Hawkins and Maboudou-Tchao, 2013]{hawkins2013}
Hawkins, D.~M. and Maboudou-Tchao, E.~M. (2013).
\newblock Smoothed linear modeling for smooth spectral data.
\newblock {\em International Journal of Spectroscopy}, 2013.

\bibitem[Li and Zhang, 2017]{li2016tensor}
Li, L. and Zhang, X. (2017).
\newblock Parsimonious tensor response regression.
\newblock {\em Journal of the American Statistical Association}, pages 1--16.

\bibitem[Ma and Zhang, 2015]{Ma2015}
Ma, Y. and Zhang, X. (2015).
\newblock A validated information criterion to determine the structural
  dimension in dimension reduction models.
\newblock {\em Biometrika}, page asv004.

\bibitem[Schott, 1994]{schott1994determining}
Schott, J.~R. (1994).
\newblock Determining the dimensionality in sliced inverse regression.
\newblock {\em Journal of the American Statistical Association},
  89(425):141--148.

\bibitem[Schott, 2013]{Schott2013}
Schott, J.~R. (2013).
\newblock On the likelihood ratio test for envelope models in multivariate
  linear regression.
\newblock {\em Biometrika}, 100(2):531--537.

\bibitem[Schwarz et~al., 1978]{bic}
Schwarz, G. et~al. (1978).
\newblock Estimating the dimension of a model.
\newblock {\em The annals of statistics}, 6(2):461--464.

\bibitem[Su and Cook, 2011]{SuCook2011partial}
Su, Z. and Cook, R.~D. (2011).
\newblock Partial envelopes for efficient estimation in multivariate linear
  regression.
\newblock {\em Biometrika}, 98(1):133--146.

\bibitem[Zeng, 2008]{Zeng2008}
Zeng, P. (2008).
\newblock Determining the dimension of the central subspace and central mean
  subspace.
\newblock {\em Biometrika}, 95(2):469--479.

\bibitem[Zhang and Li, 2017]{zhang2016tensor}
Zhang, X. and Li, L. (2017).
\newblock Tensor envelope partial least squares regression.
\newblock {\em Technometrics}, (just-accepted).

\bibitem[Zhu et~al., 2006]{zhu2006sliced}
Zhu, L., Miao, B., and Peng, H. (2006).
\newblock On sliced inverse regression with high-dimensional covariates.
\newblock {\em Journal of the American Statistical Association},
  101(474):630--643.

\bibitem[Zhu et~al., 2010]{zhu2010dimension}
Zhu, L.-P., Zhu, L.-X., and Feng, Z.-H. (2010).
\newblock Dimension reduction in regressions through cumulative slicing
  estimation.
\newblock {\em Journal of the American Statistical Association},
  105(492):1455--1466.

\bibitem[Zhu et~al., 2016]{zhu2016dimensionality}
Zhu, X., Wang, T., and Zhu, L. (2016).
\newblock Dimensionality determination: a thresholding double ridge ratio
  criterion.
\newblock {\em arXiv preprint arXiv:1608.04457}.

\bibitem[Zou and Chen, 2012]{ZouChen2012BIC}
Zou, C. and Chen, X. (2012).
\newblock On the consistency of coordinate-independent sparse estimation with
  bic.
\newblock {\em Journal of Multivariate Analysis}, 112:248--255.

\end{thebibliography}
\bibliographystyle{apalike}

\end{document}